\documentclass[prb,reprint,floatfix]{revtex4-1}
\usepackage[utf8]{inputenc}

\usepackage[british]{babel}
\usepackage{graphicx}
\graphicspath{{plot/}}
\usepackage{color}
\usepackage{amsmath,amssymb,bbold}
\usepackage{csquotes}
\usepackage{hyperref}

\renewcommand{\vec}[1]{\mathbf{#1}}
\newcommand{\dd}{\mathrm{d}}
\newcommand{\ee}{\mathrm{e}}

\newcommand{\aff}{Theoretical Physics II, University of Bayreuth, Universitätsstr. 30, 95447 Bayreuth, Germany}

\begin{document}
\title{Memory-induced motion reversal in Brownian liquids}
\author{Lucas L. Treffenstädt}
\affiliation{\aff}
\author{Matthias Schmidt}
\affiliation{\aff}
\date{8th October 2019}

\begin{abstract}
We study the Brownian dynamics of hard spheres under spatially inhomogeneous shear, using event-driven Brownian dynamics simulations and power functional theory. 
We examine density and current profiles both for steady states and for the transient dynamics after switching on and switching off an external square wave shear force field.
We find that a dense hard sphere fluid (volume fraction $\approx 0.35$) undergoes global motion reversal after switching off the shear force field.
We use power functional theory with a spatially nonlocal memory kernel to describe the superadiabatic force contributions and obtain good quantitative agreement of the theoretical results with simulation data.
The theory provides an explanation for the motion reversal: 
Internal superadiabatic nonequilibrium forces that oppose the externally driven current arise due to memory after switching off.
The effect is genuinely viscoelastic:
in steady state, viscous forces oppose the current, but they elastically generate an opposing current after switch-off.
\end{abstract}

\maketitle

\section{Introduction}
The non-equilibrium properties of hard spheres under shear have attracted considerable attention.
Rheological experiments under steady shear, e.g. using silica particles \cite{dekruif1985,marshall1990}, show non-Newtonian viscosity effects, with both shear thickening and shear thinning occurring depending on the volume fraction.
Shear thinning was observed in Brownian dynamics (BD) simulation, e.g. by Foss and Brady \cite{foss2000}.
Dhont et al. studied the distortion of the microstructure of colloids using light scattering experiments~\cite{dhont2003}.
Dhont and Nägele derived the viscoelastic response of a suspension of colloids to shear from the Smoluchowski equation~\cite{dhont1998}.
Fuchs and coworkers have developed theoretical descriptions of these effects using mode coupling theory and integration through transients \cite{cates2004,fuchs2005,fuchs2009}.
A thorough overview of the nonlinear rheology of colloidal dispersions has been given by Brader \cite{brader2010}.

Hard spheres under inhomogeneous shear exhibit a broad range of effects.
In particular, inhomogeneities in the shear rate can induce particle migration \cite{leighton1987} and thus lead to inhomogeneities in the density profile.
Examples of this mechanism are lane formation, where particles move in stacked layers separated by low density bands \cite{chakrabarti2004,waechtler2016},
and deformation of boundary density profiles of sheared systems in confinement \cite{brader2011,aerov2014,aerov2015}.
Howon et al. studied flow instabilities in inhomogeneous shear with Browninan dynamics simulations~\cite{howon2014}.

The transient behaviour in the time evolution from equilibrium to a sheared steady state and the reverse process from steady shear to equilibrium has attracted similar attention.
Reinhardt et al. \cite{reinhardt2013} studied the distortion of the pair correlation function under start-up shear.
Koumakis et al. \cite{koumakis2016} reported on stresses in the start-up phase of shearing, in particular on the dependence of the stress overshoot on the Peclet number and on the volume fraction, using both simulation and experiments with sterically stabilized PMMA spheres using confocal microscopy and rheological measurements.
Stress overshoot in start-up and cessation of shear and the connection to the microscopic fluid structure have also been studied \cite{marenne2017}.
Ackerson et al. \cite{ackerson1988} reported on solid-like ordering of nearly hard spheres under the influence of oscillatory shear.
Krüger and Brader applied dynamic density functional theory~\cite{archer2007,hopkins2010}, extended to sheared systems with a scattering kernel approach \cite{brader2011}, to study sedimentation of colloids under time-dependend shear \cite{krueger2011}, and Metz\-ger and Butler examined the time evolution of particle clusters in periodic shear \cite{metzger2012}.

Microscopic methods such as BD or molecular dynamics simulations are based on equations of motions which are instantaneous in time on the many-body level.
However, on the one-body level, nonequilibrium states are generally dependend on the history of the system.
By integrating out degrees of freedom, coarse-grained methods can be obtained, which generally have non-Markovian form, as can be shown with the Mori-Zwanzig formalism \cite{zwanzig1961,mori1965}.
There is previous work done to derive accurate memory kernels for generalised Langevin equations for Brownian dynamics.
Smith and Harris \cite{smith1990} proposed a method to approximate memory kernels and generate random forces with a given autocorrelation.
Szymczak and Cichocki \cite{szymczak2004} studied memory in the macroscopic dynamics of Brownian systems.
Bao et al. \cite{bao2005} investigated breaking of ergodicity due to memory in non-Markovian Brownian dynamics.
Recently, iterative methods have been developed to reconstruct memory kernels for generalized Langevin equations from molecular dynamics simulations by matching the force autocorrelation function or the velocity autocorrelation function between both methods \cite{lesnicki2016,jung2017}.

\begin{figure}
	\centering
	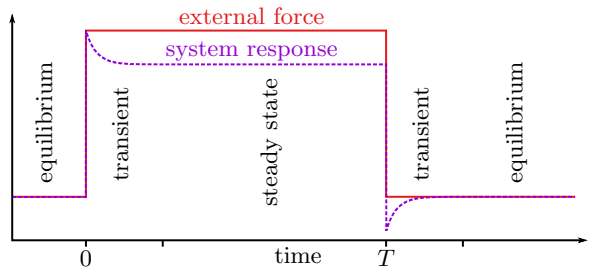
	\caption{Sketch of the time evolution of the system, external force (solid red line) and system response (dashed purple line) in arbitrary units. 
	The system was in equilibrium at negative times.
	An external force is switched on at $t=0$, and the system is monitored during the transient into a steady state as well as in the steady state itself.
	At time $t=T$, the external force is switched off and the system is observed until it has reached equilibrium again.}
	\label{fig:overview}
\end{figure}

In this paper, we examine a system of Brownian hard spheres both in steady state under temporally constant but spatially inhomogeneous shear as well as the transient dynamics after switching the driving field both on and off.
Figure~\ref{fig:overview} shows a sketch of the dynamics: Starting in a well-defined equilibrium state, a shear force field is switched on.
The system needs some time to relax into a steady state.
Then, the shear force field is switched off, and the system relaxes back into equilibrium.
We report in particular on the shape of the current profile in the steady state under the influence of a square wave shear profile.
This particular form of shear is well suited to show and examine nonlocal effects, since small regions of extreme shear rate alternate with large regions of low shear rate.
We find that the transition in the current field between opposite flow directions is non-monotonic.
After switching-off of the driving shear force, the current field reverses globally before settling into equilibrium.

We employ state-of-the-art event driven Brownian dynamics simulations \cite{scala2007}, which solve the problem of infinite gradients in the hard sphere interaction potential by evolving the system continuously with ballistic motion between BD timesteps.
At fixed timesteps, the velocities of the particles are randomised according to a Maxwell distribution.

In addition to observations in simulation, the system is examined in the framework of power functional theory (PFT)\cite{schmidt2013}, which describes the full non-equilibrium dynamics of many-particle systems, beyond the adiabatic approximation made in dynamical density functional theory (DDFT).
DDFT is an extension of equilibrium density functional theory (DFT) to nonequilibrium systems \cite{marconi1999,archer2004},
which approximates the time evolution of the system through a series of adiabatic states, where the internal forces can be calculated from an equivalent equilibrium system with matching instantaneous density~\cite{fortini2014}.
However, this approximation leads to shortcomings, such as underestimation of relaxation times \cite{marconi1999}.
There have been attempts to correct these shortcomings via empirical corrections, see e.g.~\cite{stopper2015,stopper2015b}.

Superadiabatic forces were shown to occur in a variety of systems, such as Gaussian core particles \cite{stuhlmueller2018}, hard spheres \cite{heras2018} and active Brownian particles \cite{krinninger2019}.
We extend here an approximation for superadiabatic forces for Brownian hard spheres, presented recently by de las Heras and Schmidt \cite{heras2018} by introducing a diffusing memory kernel.
This approximation derives forces from the gradient of the velocity field.
The free parameters in this model -- memory time, memory diffusion constant, and overall memory strength -- are determined using a least-squares fit to BD simulation data.

This paper is organised as follows:
In section II, we introduce the considered system and our PFT approach.
Section III contains implementation details for the BD simulations.
Sections IV and V cover results in steady state and during transients, respectively.
We draw conclusions and provide an outlook in section VI.

\section{System and Theory}
We consider a fluid of $N$ monodisperse hard spheres with diameter $\sigma$.
The system has planar geometry with Cartesian coordinates $\vec r = (x,y,z)$ and we take $\sigma$ as the unit of length.
Isotropy is broken by an external shear force field
\begin{equation}
	\label{eq:fext}
	\vec{f_\mathrm{ext}}(\vec{r}) = f_\mathrm{ext}(x)\vec{\hat{e}}_\mathrm{z}\text{,}
\end{equation}
where $\vec{\hat{e}}_\mathrm{z}$ is the unit vector in the $z$-direction and $f_\mathrm{ext} = \left|\vec f_\mathrm{exc}\right|$ is the modulus of the force field.
Since the intrinsic dynamics are diffusive, we choose as the unit of time the diffusion time $\tau = \sigma^2/D$ with diffusion constant $D = k_\mathrm{B}T/\gamma$, where $k_\mathrm{B}$ is the Boltzmann constant, $T$ indicates the absolute temperature, and $\gamma$ is the friction constant against the implicit solvent.

The particle positions $\vec r_1,\ldots,\vec r_N \equiv \vec r^N$ evolve in time according to the Langevin equation of motion
\begin{equation}
	\label{eq:langevin}
	\gamma \dot{\vec{r}_i}(t) = \vec f_{\mathrm{int},i}(\vec r^N) + \vec f_\mathrm{ext}(\vec r_i,t) + \sqrt{2\gamma k_\mathrm{B}T}\vec R_i(t)
\end{equation}
where $\vec f_{\mathrm{int},i} = -\nabla_i u(\vec r^N)$ is the internal force that all other particles exert on particle $i$ due to the interaction potential $u(\vec r^N)$ and $\vec R_i(t)$ is a delta-correlated Gaussian random white noise with $\left<\vec R_i(t)\right> = 0$ and $\left<\vec R_i(t) \vec R_j(t')\right> = \delta(t-t')\delta_{ij}\mathbb{1}$, where $\delta(\cdot)$ is the Dirac distribution, $\delta_{ij}$ indicates the Kronecker delta, and $\mathbb{1}$ is the $3\times3$ unit matrix.

The one-body density distribution is defined as
\begin{equation}
	\label{eq:density}
	\rho(\vec r, t) = \left<\sum\limits_i \delta(\vec r - \vec r_i)\right>\text{,}
\end{equation}
where $\left<\cdot\right>$ indicates an average over the noise and over initial microstates.
The one-body current distribution is defined as 
\begin{equation}
	\label{eq:current}
	\vec J(\vec r, t) = \left< \sum\limits_i \delta(\vec r - \vec r_i)\vec v_i(t) \right>\text{,}
\end{equation}
where, in a numerical simulation, $\vec v_i$ must be calculated with a finite difference centred at time $t$ \cite{heras2019}.
The velocity field $\vec v(\vec r,t)$ is defined as
\begin{equation}
	\label{eq:velocity}
	\vec v(\vec r,t) = \dfrac{\vec J(\vec r,t)}{\rho(\vec r,t)}\text{.}
\end{equation}
The dynamics of \eqref{eq:density} and \eqref{eq:current} can be expressed as
\begin{align}
	\label{eq:motionj}
	\gamma \vec v(\vec r,t) &= \vec f_\mathrm{int} + \vec f_\mathrm{ext} -\mathrm{k_B}T\nabla \ln \rho\text{,}\\
	\label{eq:continuity}
	\frac{\partial}{\partial t}\rho(\vec r,t) &= -\nabla \cdot \vec J(\vec r,t)\text{,}
\end{align}
with total internal force field
\begin{equation}
	\vec f_\mathrm{int}(\vec r,t) = \dfrac{1}{\rho}\left< \sum\limits_i \delta(\vec r - \vec r_i)\vec f_{\mathrm{int},i} \right> \text{.}
\end{equation}
This internal force field can be split into two parts according to
\begin{equation}
	\vec f_\mathrm{int} = \vec f_\mathrm{ad} + \vec f_\mathrm{sup}\text{,}
\end{equation}
with the adiabatic force $\vec f_\mathrm{ad}$ and the superadiabatic force $\vec f_\mathrm{sup}$\cite{schmidt2013,fortini2014}.
The adiabatic force is defined as the internal force acting in a constructed equilibrium system with an external potential $V_\mathrm{ad}(\vec r)$ chosen such that the equilibrium density matches the instantaneous density $\rho(\vec r,t)$.
The underlying map from the equilibrium density distribution to the external potential $V_\mathrm{ad}$ has been shown by Evans \cite{evans1979} and Mermin \cite{mermin1965}.
Thus, $\vec f_\mathrm{ad}$ depends only on the density at time $t$.
The superadiabatic force field, in contrast, depends in general on the history of both $\rho(\vec r,t^\prime)$ and $\vec J(\vec r,t^\prime)$ for $t^\prime \le t$, making \eqref{eq:motionj} in general an implicit equation.
This distinction physically defines the splitting of internal forces.
Superadiabatic forces can be measured in particle-based simulations~\cite{fortini2014}.

Power functional theory is based on the free power functional $R_t[\rho,\vec J]$, which captures in a formally exact way the full many-body dynamics.
$R_t$ generates via a minimisation principle
\begin{equation}
	\label{eq:euler}
	\dfrac{\delta R_t}{\delta \vec J(\vec r,t)} = 0\text{ (min),}
\end{equation}
an Euler-Lagrange equation of motion \eqref{eq:motionj} for the current, given a density profile at fixed time $t$ and the history $\rho(\vec r,\tilde{t}),\vec J(\vec r,\tilde{t})$ for $\tilde{t} < t$.
The resulting current at time $t$ can then be used in conjuction with the continuity equation~\eqref{eq:continuity} to evolve the density in time.

The power functional $R_t[\rho,\vec{J}]$ for a many-body system splits into
\begin{equation}
	R_t[\rho,\vec{J}] = \dot{F} + P_t - X_t\text{,}
\end{equation}
where
\begin{equation}
	\dot{F}[\rho] = \int \dd \vec r\; \vec{J}\cdot\nabla\dfrac{\delta F[\rho]}{\delta \rho}\text{,}
\end{equation}
is the time derivative of the intrinsic Helmholtz free energy functional $F[\rho]$ of equilibrium DFT.
$F[\rho]$ can be split into an ideal part $F_\mathrm{id}[\rho]$, which contains contributions to the free energy from the ideal gas, and the excess free energy $F_\mathrm{exc}[\rho]$, which depends on the particle interactions.
For the excess free energy functional, we choose the well-known Rosenfeld functional \cite{rosenfeld1989}.
The negative functional derivative of $\dot{F}_\mathrm{exc}$ with respect to $\vec J$ produces the adiabatic forces, which only depend on the density $\rho$ at time $t$.
Hence
\begin{equation}
	\vec f_\mathrm{ad} = -\dfrac{\delta \dot{F}_\mathrm{exc}}{\delta \vec J} = - \nabla \dfrac{\delta F_\mathrm{exc}}{\delta \rho}\text{.}
\end{equation}

The external power $X_t$ depends on the external force field $\vec{f}_\mathrm{ext}$, as well as the time derivative $\dot{V}_\mathrm{ext}$ of the external potential, should it be time-dependend.
The functional has the form
\begin{equation}
	\label{eq:xt}
	X_t = \int \dd \vec r \; \left[\vec{J}\cdot \vec{f}_\mathrm{ext}(\vec r,t) - \rho \dot{V}_\mathrm{ext}\right]\text{.}
\end{equation}
Here, $\dot{V}_\mathrm{ext} = 0$ and the external force field is a shearing force (cf. eq.~\eqref{eq:fext}).
Additionally, we employ a temporally constant conservative force field $\vec f_\mathrm{c} = -\nabla V_\mathrm{ext}$ to induce particle migration effects.

Finally, $P_t$ can be split into an ideal and an excess (over-ideal) part
\begin{equation}
	P_t = P_t^\mathrm{id} + P_t^\mathrm{exc}\text{,}
\end{equation}
with the ideal dissipation functional
\begin{equation}
	P_t^\mathrm{id} = \gamma \int \dd \vec r \dfrac{\vec{J}^2}{2\rho}\text{,}
\end{equation}
which is local in time and space, as is appropriate for free diffusion.

$P_t^\mathrm{exc}$ contains all superadiabatic effects and is, in general, nonlocal in both space and time via causal history dependence.
It generates the superadiabatic forces via
\begin{equation}
	\vec f_\mathrm{sup} = -\dfrac{\delta P_t^\mathrm{exc}}{\delta \vec J}\text{.}
\end{equation}
$P_t^\mathrm{exc}$ is specific to the type of interparticle interaction potential and must in general be approximated.
This status is very similar to that of the excess free energy functional $F_\mathrm{exc}[\rho]$ in equilibrium DFT.
Here, we choose the generic velocity gradient approximation \cite{heras2018}

\begin{widetext}
\begin{equation}
	\label{eqn:pexc}
	P_t^\mathrm{exc} = \dfrac{\gamma}{2} \int\dd \vec r \int\dd \vec r' \int\limits_{-\infty}^t \dd t' \rho(\vec r,t)\left[\eta(\nabla\times\vec v)\cdot(\nabla'\times\vec v') + 
	\zeta(\nabla\cdot\vec v)(\nabla'\cdot\vec v')\right]\rho(\vec r',t')K(\vec r-\vec r',t-t')\text{,}
\end{equation}
\end{widetext}
where $\vec v = \vec v(\vec r,t)$ (and $\vec v' = \vec v(\vec r',t')$) is the velocity field as defined in \eqref{eq:velocity}, and $\nabla'$ is the derivative with respect to $\vec r'$.
$P_t^\mathrm{exc}$ depends not only on the instantaneous density and velocity fields, but also on the history of the system, and it is non-local in space.
$\zeta$ and $\eta$ is the volume viscosity and the shear viscosity, respectively.
The particular coupling to the history is governed by the memory kernel $K(\vec r-\vec r',t-t')$, which is normalised  such that $\int \dd \vec r \int \dd \vec r^\prime \int \dd t^\prime K = 1$.

We examine two different functional forms of $K$.
The simpler of the two is local in space:
\begin{equation}
	\label{eq:KL}
	K_\mathrm{L}(\vec r,t) = \delta(\vec r){\tau_\mathrm{M}}^{-1}\exp\left(-t/\tau_\mathrm{M}\right)\Theta(t)\text{,}
\end{equation}
with memory time $\tau_\mathrm{M}$ and the Heaviside step function $\Theta(\cdot)$.
We expect this form to perform well in cases of mild shear rates, where it is well suited to explore time-dependend behaviour in isolation from spatial effects.

The second version is spatially non-local and based on the idea that interactions between distant particles propagate according to the underlying microscopic dynamics, which are diffusive.
We introduce a corresponding memory diffusion constant $D_\mathrm{M}$.
The memory kernel takes the form
\begin{equation}
	\label{eq:KD}
	K_\mathrm{D}(\vec r,t) = \Theta(t)\frac{1}{\tau_\mathrm{M}}\ee^{-\dfrac{t}{\tau_\mathrm{M}}}\left(4\pi t D_\mathrm{M}\right)^{-3/2}\ee^{-\dfrac{\vec r^2}{4 tD_\mathrm{M}}}\text{,}
\end{equation}
with memory time $\tau_\mathrm{M}$ as before.
We shall call this form the diffusing memory kernel, since the spatial part corresponds to a simple diffusion process.
The constants $\tau_\mathrm{M}$ and $D_\mathrm{M}$ are treated as free parameters.

The timescale $\tau_\mathrm{M}$ controls the exponential decay of the memory effect.
$D_\mathrm{M}$ has the units of a square length per time and controls how fast information from a point $\vec r'$ can reach the point $\vec r$.
This corresponds to a diffusion process.
In the limit of $t' \rightarrow t$, the spatial part of $K_\mathrm{D}$ approaches the Dirac delta distribution.
Therefore, there are no instantaneous non-local interactions in this model.

In steady state, density and current do not depend on time, i.e. $\rho(\vec r,t) = \rho_\mathrm{s}(\vec r)$ and $\vec{J}(\vec r,t) = \vec{J}_\mathrm{s}(\vec r)$ with $\nabla \cdot \vec J_\mathrm{s} = 0$.
In this case, the time integral in~\eqref{eqn:pexc} acts only on $K$ and, as $K$ is known from \eqref{eq:KL} or \eqref{eq:KD}, can be carried out explicitly.
The respective results for both kernels are

\begin{align}
	K_\mathrm{L}^\mathrm{S} = \int\limits_{-\infty}^t &K_\mathrm{L}(\vec r-\vec r',t-t')\dd t' = \delta(\vec r-\vec r')\text{,} \\
\begin{split}
\label{eq:kd_steady}
	K_\mathrm{D}^\mathrm{S} = \int\limits_{-\infty}^t &K_\mathrm{D}(\vec r-\vec r',t-t')\dd t' \\
	&=\dfrac{1}{4\pi\tau_\mathrm{M}D_\mathrm{M}|\vec r-\vec r'|}\exp\left(-\dfrac{|\vec r-\vec r'|}{\sqrt{\tau_\mathrm{M}D_\mathrm{M}}}\right)\text{.}
\end{split}
\end{align}

Thus, $K_\mathrm{L}^\mathrm{S}$ does not depend on the parameter $\tau_\mathrm{M}$ and
$K_\mathrm{D}^\mathrm{S}$ depends only on a new length scale
\begin{equation}
	\label{eq:sigma_m}
	\sigma_\mathrm{M} = \sqrt{\tau_\mathrm{M}D_\mathrm{M}}\text{,}
\end{equation}
which can be interpreted as an effective interaction length in steady state.
The parameters $\tau_\mathrm{M}$ and $D_\mathrm{M}$ can therefore not be independently determined from measurements of one-body quantities in steady state.
However, one can determine the value of $\sigma_\mathrm{M}$.
In steady state, it is less computationally intensive to obtain accurate density and current profiles from particle simulations, so $\sigma_\mathrm{M}$ can be determined with high accuracy.
Knowledge of $\sigma_\mathrm{M}$ then reduces the number of free parameters to be determined with measurements in the full time evolution.

For the given system, all integrals in $y$ and $z$ in $R_t[\rho,\vec J]$ can be explicitly carried out, since density and current are by construction homogeneous in these directions.
Thus, the current only depends on one space coordinate $x$ and time $t$: $\vec J(x,t) = J_x(x,t) \hat{\vec e}_x + J_z(x,t) \hat{\vec e}_z$,
where $J_x$ is the current in gradient direction $\hat{\vec{e}}_x$, and $J_z$ is the current in flow direction $\hat{\vec{e}}_z$.

It should be noted that the form of $P_t^\mathrm{exc}$ applied here contains no coupling between the flow direction and the gradient direction of $\vec{J}$.
Therefore, a system with an initially homogeneous density and no external force acting in the $x$ direction will always remain homogeneous in this approximation, whereas in reality, structural migration forces occur.
$P_t^\mathrm{exc}$ can be extended to include these effects \cite{stuhlmueller2018}, but that is beyond the scope of this work.
Instead, we impose the density profile $\rho_\mathrm{BD}$ obtained in BD simulations via an external potential $V_\mathrm{ext}(x)$, chosen so that $\rho_\mathrm{BD}(x)$ is the equilibrium density in the potential.

We numerically minimize $R_t[\rho,\vec{J}]$ for a given $\rho(x,t)$ at time $t$ using a generic nonlinear numerical optimiser \cite{nlopt}, thus solving the Euler-Lagrange-equation~\eqref{eq:euler} and obtaining $\vec{J}(x,t)$.
Using the continuity equation \eqref{eq:continuity}, we numerically evolve $\rho$ in time, i.e. proceed by one time step $\Delta t$ and repeat the procedure.

Then, we compare results for $\rho(x,t)$ and $\vec{J}(x,t)$ calculated with PFT to results for the same quantities sampled in BD simulations.
The free parameters in $P_t^\mathrm{exc}$ can be determined via a least-squares fit with an appropriate observable.
We choose here the velocity field, sampled at fixed time intervals during transients, or averaged over multiple simulation snapshots in the case of steady state.
In principle, other observables could be used.
We start from reasonable estimates and use a nonlinear numerical optimiser \cite{nlopt} with a derivative-free optimisation routine \cite{powell1994} to obtain estimates of the free parameters.

\section{Brownian Dynamics Simulations}
We employ event-driven Brownian Dynamics simulations \cite{scala2007} to integrate the Langevin equation~\eqref{eq:langevin} and obtain particle trajectories.
We use $N = 1090$ particles in a simulation box of size $10\times10\times15~\sigma^3$ with periodic boundary conditions in all directions.
By choosing a strongly inhomogeneous shear force field, we expect to clearly showcase the importance of nonlocal interactions.
Our choice of a field that is periodic in $x$ relieves us from the need for Lees-Edwards boundary conditions~\cite{lees1972}, which are commonly used for periodic systems with constant shear rate.

We calculate one-body quantities such as density and current by averaging over many-body trajectories.

We obtain the steady state current and density profile by averaging $10^6$ trajectory samples from a runlength of $10^3 \tau$ after an initial relaxation period of $2\tau$.
For start-up dynamics, the system is simulated in equilibrium for an initial $0.1\tau$, after which shear is switched on and the system is evolved for a further $0.4\tau$.
Dynamics after switch-off are simulated initially for $1.5\tau$ under shear, after which the shear force is switched off and the system is evolved for a further $0.1\tau$.
In our experience, this protocol is sufficient to ensure that a steady state has been reached, given our parameters.
Time-dependent current and density profiles for the dynamics in full non-equilibrium are then calculated per timestep via an average over $10^4$ realisations.

\section{Steady State}
\begin{figure}
	\centering
	\input{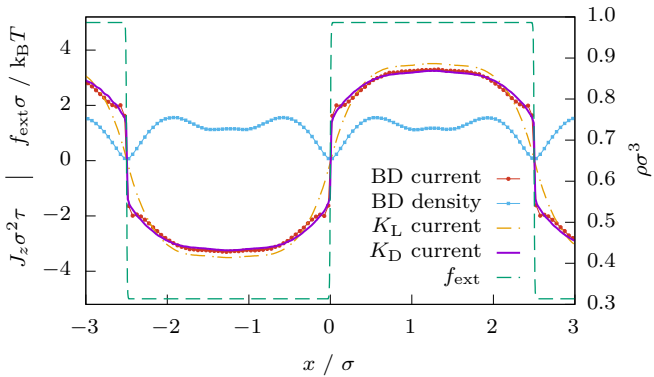}
	\caption{Steady state current $J_z(x)$ and density profile $\rho(x)$ in BD simulation under a square wave shear force $f_\mathrm{ext}(x)$ acting along $\hat{\vec e}_z$. The plot shows only part of the simulation box.}
	\label{fig:steadystate}
\end{figure}

We subject our system to a strongly inhomogeneous, but steady, square wave external force in $z$ direction with an amplitude of $5\frac{\mathrm{k_B}T}{\sigma}$ and a period of $5\sigma$ (see figure~\ref{fig:steadystate}).
After a short time $\approx 10^{-1}\tau$, the system settles into a steady state ($\partial_t \rho = \partial_t \vec J = 0$).

This state has some interesting properties (Figure~\ref{fig:steadystate}, current and density profile in steady state).
The BD results indicate that, even though no external force is acting in the $x$ direction, the density profile becomes inhomogeneous.
This effect is driven purely by superadiabatic forces and is thus a true nonequilibrium effect.
Phenomenological approaches to incorporate such forces into DDFT have been proposed in~\cite{brader2010,krueger2011}.
Stuhlmüller et al.~\cite{stuhlmueller2018} have studied shear induced particle migration in a system of Gaussian core particles with PFT.
In our PFT calculations, we impose the inhomogeneous density sampled in BD with a temporally constant external potential $V_\mathrm{ext}(x)$.

The harsh spatial step in the driving force field is reflected in the current profile: The current reverses its orientation in a region smaller than $\sigma/10$.
Inside the regions of near-constant force, instead of a monotonic approach to the maximum, the current profile displays an oscillation close to the edge.
The occurence of this effect suggests a complex nonlocal interaction, supporting our corresponding approach in PFT.

Using a least-squares fit of the PFT velocity profile to the BD velocity profile, we can obtain values for $\eta$ in $K_\mathrm{L}$ and $\sigma_\mathrm{M}$ in $K_\mathrm{D}$~\eqref{eq:kd_steady},\eqref{eq:sigma_m}.
Figure~\ref{fig:steadystate} shows the resulting velocity profiles from PFT.
In the given case, we obtain $\sigma_\mathrm{M} \approx \sigma/3$, which is close to the sphere radius.
The effective interaction in steady state is therefore quite short-ranged.

While not perfect, the agreement between BD and PFT is much better for the diffusing memory form~\eqref{eq:KD} than it is for the local form~\eqref{eq:KL}.
Perhaps contrary to intuition, the profile obtained from the local memory model is smoother and does not represent the jump in the current profile that is observed in BD.
The reason for this becomes clear when considering the effect of the spatial nonlocality of $K_\mathrm{D}$.

The velocity gradient $\partial_{x} v_\mathrm{z}$ has a large spike at the jump of the velocity itself.
In the local memory model, this spike contributes evenly for every point in the history of the system.
In the diffusing model, it is smoothed out by the integral over $x'$ for times $t' < t$.
The penalty for a jump in the velocity is thus much lower in the diffusing model.

Since $P_t^\mathrm{exc}$ depends only on inter-particle interactions and not on external forces, and should be translationally invariant, no spatially local memory kernel can accurately represent this feature in the velocity profile, no matter how complex the temporal behaviour.
In other words, spatial nonlocality is not only the most general form of memory, but it is required for the correct description of strong inhomogeneities within the velocity gradient approach.

\section{Transient Dynamics}
\begin{figure}
	\centering
	\input{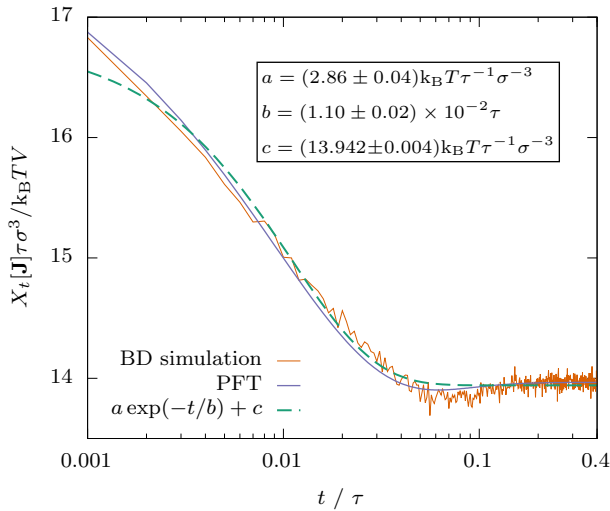}
	\caption{Time evolution (on a logarithmic scale) of the scaled mean external power $X_t$ per volume after switching on the shear force at $t=0$, from BD simulation and PFT, together with a least-squares fit of a simple exponential decay $a\exp(-t/b)+c$.}
	\label{fig:evolution-on}
\end{figure}
\begin{figure*}
	\centering
	\input{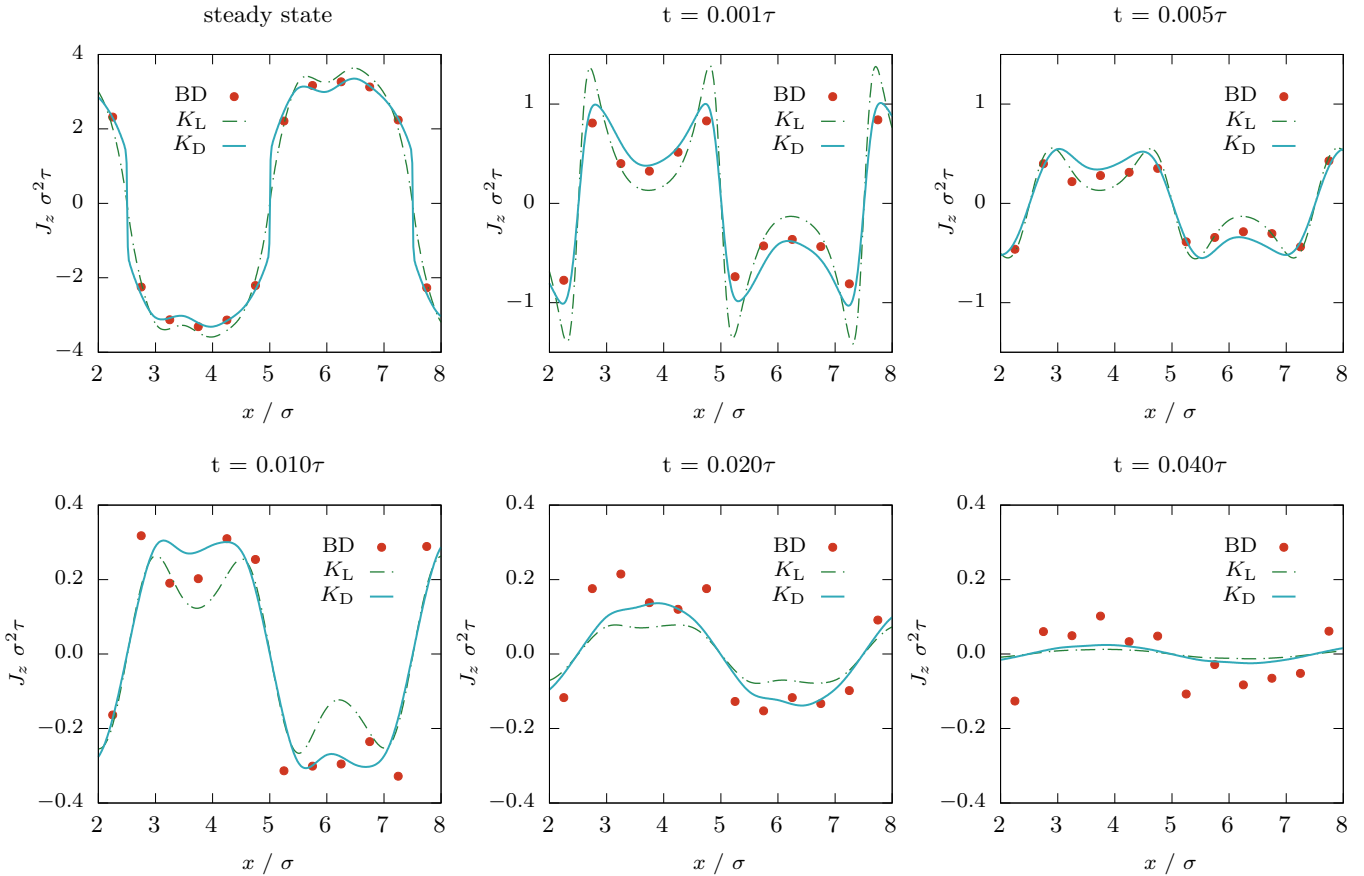}
	\caption{Time evolution of the scaled current profile $J_x \sigma^2 \tau$ as a function of $x \ \sigma$ after switching off the shear force at $t=0$, obtained from BD simulation (symbols), and PFT with local memory kernel $K_\mathrm{L}$ (dashed line) and diffusing memory kernel $K_\mathrm{D}$ (solid line). The sign of the current flips globally after the shear force is switched off, followed by a decay into equilibrium. The diffusing memory model in PFT significantly outperforms the local approach in representing the BD current profiles.}
	\label{fig:evolution-off}
\end{figure*}

We next investigate the transient dynamics into and out of the steady state.
First, we address the transient going from equilibrium to a sheared system.
The external shear force field is the same as above.
It is switched on instantaneously at $t=0$ and switched off again at a later time.
The system responds instantly to the external force, with an instantaneous current profile that has \emph{larger magnitude} than the steady state current.
This instant response to the external force is consistent with the fact that inertia is neglected in overdamped BD.
Then, viscosity slows the system down into the steady state (see figure~\ref{fig:evolution-on}).
This process takes finite time, because the particles need to traverse, on average, the mean free path $\lambda = (\sqrt{2}\pi\rho\sigma^2)^{-1}$ between collisions~\cite{hansen2013}, which takes, with diffusive dynamics, about $\lambda^2/D$, which is $\approx 0.1\tau$ in our system.

The decay of the mean external power~\eqref{eq:xt} into the steady state can be described reasonably well with an exponential decay model $X_t/V = a\exp(-t/b)+c$, where $a$ and $c$ are parameters with the dimension of power per volume, and $b$ is a parameter with the dimension of a time.
Physically, $c$ can be identified as the mean steady state external power density, $a$ as the initial mean super-steady external power density, and $b$ as the decay time.
However, the decay shows features beyond a simple exponential, which are captured by PFT with diffusing memory kernel.
For times $0.05 \le t/\tau \le 0.1$, the external power dips below the plateau value, and the initial decay is steeper than exponential.

Next, we explore the dynamics after switching off the shear force (see figure~\ref{fig:evolution-off}).
Surprisingly, the current does not relax monotonically into equilibrium, but rather undergoes a global reversal first, and then smoothly equilibrates.
This remarkable result has perhaps been hinted at by Krüger and Brader \cite{krueger2011}, who report "If the shear field is suddenly switched off, we find that the equilibration dynamics show an interesting symmetry with that following switch on[...]."
Other than that, to the best of our knowledge, this effect has not been reported in the literature.

Using the time-dependend velocity field measured in BD for switch-on and switch-off, we can determine the remaining free parameters in $K_\mathrm{L}$ and $K_\mathrm{D}$.
The exponential decay in both $K_\mathrm{L}$ and $K_\mathrm{D}$ is one of the most simple forms of memory.
Starting from an initial equilibrium state, the memory integral at time $t=0$ vanishes, because the velocity gradient vanishes at negative times.
Therefore, the superadiabatic force field is also zero just after switching on, and the current is directly proportional to the driving force. 

Memory then slowly builds up, with a dynamical behaviour that is governed in our approximation by the memory time $\tau_\mathrm{M}$.
We obtain memory times of roughly $\tau_\mathrm{M} = 0.02\tau$ for switch-on.
The superadiabatic forces oppose the current, slowing the system into a steady state.

The steady state is truely reached once the current has not changed over a few memory times $\tau_\mathrm{M}$ and thus the memory integral no longer changes.
Then, the driving force can be switched off and the transient back into equilibrium can be observed.
PFT accurately predicts the motion reversal observed in BD and provides an explanation:
In the steady state, the force balance (cf. eq.~\eqref{eq:motionj})
includes adiabatic forces $\vec f_\mathrm{ad}$, external forces $\vec f_\mathrm{ext}$ and superadiabatic forces $\vec f_\mathrm{sup}$.
In the direction of shear, $\vec f_\mathrm{ad}\cdot \vec{\hat{e}}_z = 0$ because of the homogeneity of $\rho$ in $z$.
As we know, $\vec{f}_\mathrm{sup}$ is opposed to the external force.
With $\vec{f}_\mathrm{ext} = 0$ after switching-off, the superadiabatic excess forces still remain, because they arise from the memory integral.
Thus, the superadiabatic forces become driving forces with an opposed direction of motion.
The system returns to equilibrium only after the memory has cleared.
The memory time obtained here is roughly $\tau_\mathrm{M} = 0.01\tau$.
While the decay of the counter-current is well described by the exponential decay memory model up until $t \approx 0.02\tau$, it seems to overestimate the rate of relaxation for later times (see figure~\ref{fig:evolution-off}).

\section{Conclusion}
We have studied the Brownian hard sphere fluid under inhomogeneous, time-dependent shear with BD simulations and PFT.
In steady state, under strongly inhomogeneous shear, spatially nonlocal memory shapes the current profile in ways spatially local memory cannot.
Non-local memory is therefore required to describe general external forces acting on the fluid with a true separation of intrinsic and extrinsic effects.
Exponential memory is an adequate and simple approximation that well describes nonequilibrium dynamics after switching (on and off) of an external field.
The effect of motion reversal after switch-off is surprising if thought about in a microscopic picture, but has a straightforward explanation in PFT:
Slowing memory forces in steady state become driving forces after the shear force has been switched off.
The rigorous framework of PFT is therefore an appropriate tool to gain insight into the behaviour of the Brownian hard-sphere fluid.

Non-local memory could be a relevant factor in the study of inhomogeneous colloidal systems such as colloids undergoing capillary collapse at an interface~\cite{bleibel2011}.
We are also interested to investigate the effect of the approximation presented here on the bulk dynamics of hard spheres, such as the van Hove correlation function, which has been studied recently experimentally and with DDFT~\cite{stopper2018}.
To this end, we plan to employ PFT in the dynamic test particle limit~\cite{archer2007,hopkins2010,brader2015}.

The excess superadiabatic functional can be further developed in two directions:
Spatially, structural forces can be incorporated with higher orders of the velocity gradient.
The diffusing nonlocality provides good results, but has free parameters that need to be tuned by BD simulation or other benchmarks.
Instead, they might be derived from the particle interaction, perhaps based on fundamental measures to allow for a deeper physical interpretation.
Temporally, the exponential decay model could be improved.
Research on memory in molecular dynamics provides a jumping-off point \cite{lesnicki2016}.
Recently, Jung et al \cite{jung2017} presented a method to obtain memory kernels that could be adapted to our approach.

Finally, we expect the current reversal effect, presented here for Brownian hard spheres, to re reproducible in an experimental realisation.

\section*{Conflicts of Interest}
There are no conflicts of interest to declare.

\section*{Acknowledgements}
We thank Daniel de las Heras for useful comments. This work is supported by the German Research Foundation (DFG) via SCHM 2632/1-1.

\bibliography{citations}
\end{document}